\documentclass[12pt]{article}

\usepackage[utf8]{inputenc}
\usepackage[T1]{fontenc}
\usepackage{scicite}
\usepackage{cite}
\usepackage{placeins}
\usepackage{times}
\usepackage{graphicx}
\usepackage{color}
\usepackage{setspace} 
\usepackage{amsmath}

\usepackage{verbatim}

\newenvironment{sciabstract}{%
\begin{quote} \bf}
{\end{quote}}

\title{Real-time imaging of acoustic waves in bulk materials with X-ray microscopy} 

\author
{Theodor S. Holstad,$^{1}$ Leora E. Dresselhaus-Marais,$^{2,3,4}$ \\ Trygve Magnus R\ae der,$^{1}$ Bernard Kozioziemski,$^{4}$   
 \\ Tim van Driel,$^{3}$  Matthew Seaberg,$^{3}$ \\ Eric Folsom,$^{4}$ Jon H. Eggert$^{4}$ \\
 Erik Bergb\"ack Knudsen,$^{1}$ Martin Meedom Nielsen,$^{1}$ \\Hugh Simons,$^{1}$    Kristoffer Haldrup,$^{1}$ \\ Henning Friis Poulsen$^{1\ast}$
\\ \\
\normalsize{$^{1}$Department of Physics, Technical University of Denmark; Anker Engelunds Vej 101,} \\ \normalsize{2800 Kongens Lyngby, Denmark}\\
\normalsize{$^{2}$Department of Materials Science \& Engineering, Stanford University;} \\
\normalsize{496 Lomita Mall, Stanford, CA 94305, USA}\\
\normalsize{$^{3}$SLAC National Accelerator Laboratory; 2575 Sand Hill Road,}\\ 
\normalsize{Menlo Park, CA 94025-7015, USA}\\
\normalsize{$^{4}$Physics Division, Lawrence Livermore National Laboratory; 7000 East Ave.,}\\ \normalsize{Livermore, CA 94550-9234, USA}
\\
\normalsize{$^\ast$Henning Friis Poulsen; E-mail:  hfpo@fysik.dtu.dk.}
}

\date{}

\begin{document} 


\baselineskip24pt


\maketitle 

\begin{sciabstract}

Materials modelling and processing require experiments to visualize and quantify how external excitations drive the evolution of deep subsurface structure and defects that determine properties. Today, 3D movies with $\sim$100-nm resolution of crystalline structure are regularly acquired in minutes to hours using X-ray diffraction based imaging. We present an X-ray microscope that improves this time resolution to $<$100 femtoseconds, with images attainable even from a single X-ray pulse. Using this, we resolve the propagation of 18-km/s acoustic waves in mm-sized diamond crystals, and demonstrate how mechanical energy thermalizes from picosecond to microsecond timescales. Our approach unlocks a vast range of new experiments of materials phenomena with intricate structural dynamics at ultrafast timescales. 
\end{sciabstract}


Understanding the structural dynamics of crystalline solids is a key aspect of materials science, geo-science  and solid state physics. However, the structure of many materials is complex, exhibiting dynamics across multiple length- and time-scales simultaneously. \emph{In situ} tools for visualization of such multiscale dynamics is in general lacking. As a consequence, materials models and simulations have suffered from a lack of input from experiments, and in many cases exhibit poor prediction capabilities. In particular, this is the case for structural materials, such as most metals, ceramics, rocks, and bone, where the structure is organized hierarchically in grains, domains and defects, and competing interactions take place on length scales from nanometres to centimeters \cite{Cahn1991}. For such materials, experimental techniques capable of characterizing samples that are tens or hundreds of micrometers thick are required to build reliable models. As a result, optical and electron microscopy and other probes of the (near) surface are mainly relevant for \emph{post mortem} characterization. 

To record movies of the evolution of phase transitions, grain-boundary or domain motion, and crystalline defects/strains within mm thick specimens, X-ray diffraction based imaging methods have been developed. Exploiting the brightness and penetration power of synchrotron X-ray sources, modalities include 3D X-ray Diffraction  \cite{Lauridsen2001,  Suter2006},
Diffraction-Contrast Tomography \cite{Johnson2008},  
Differential-Aperture X-ray Structural Microscopy \cite{Larson2002, Ice2011}
and Dark-Field X-ray Microscopy, DFXM \cite{Simons2015,Poulsen2018}.
With a spatial resolution down to 100 nm, these methods have been used \emph{e.g.} to reveal underlying mechanisms in nucleation and growth phenomena  \cite{Schmidt2004, Zhang2020}, in plastic deformation \cite{Margulies2001}, in fracture \cite{King2008}, in phase transformations \cite{Simons2018}, in dislocation dynamics \cite{Dresselhaus-Marais2021} and in the complex mechanics of bone \cite{Langer2016}.  

Currently, a main limitation of these methods is the time resolution: the limited brightness of the source requires acquisitions over milliseconds to seconds for each single image, making spatially resolved maps require minutes to hours for different sample volumes and imaging modalities. In contrast, many diffusive processes occur on timescales of microseconds, while numerous processes like Martensitic phase transitions, charge-density waves and thermal transport occur even faster (in pico- to nanoseconds).

In this work we demonstrate X-ray imaging within the bulk of a material that captures structural dynamics in <100 femtosecond snapshots. Specifically, we visualise the propagation of acoustic waves in a diamond single crystal as it
shifts from a mechanical “impulse” into a thermal bath. By applying DFXM at an X-ray Free Electron Laser (XFEL) our movies visualize and quantify the ultrafast propagation of the sound waves and resolve their interactions with surfaces, converting energy into transverse modes that ``ring down'' the energy over picosecond through microsecond timescales. We include coupled thermomechanical and X-ray optics simulations to illustrate the generation and imaging technique of acoustic waves in this setup \cite{Holstad2022}.

\begin{figure}[t!]
\centering
  \centering
  \includegraphics[width=.98\linewidth]{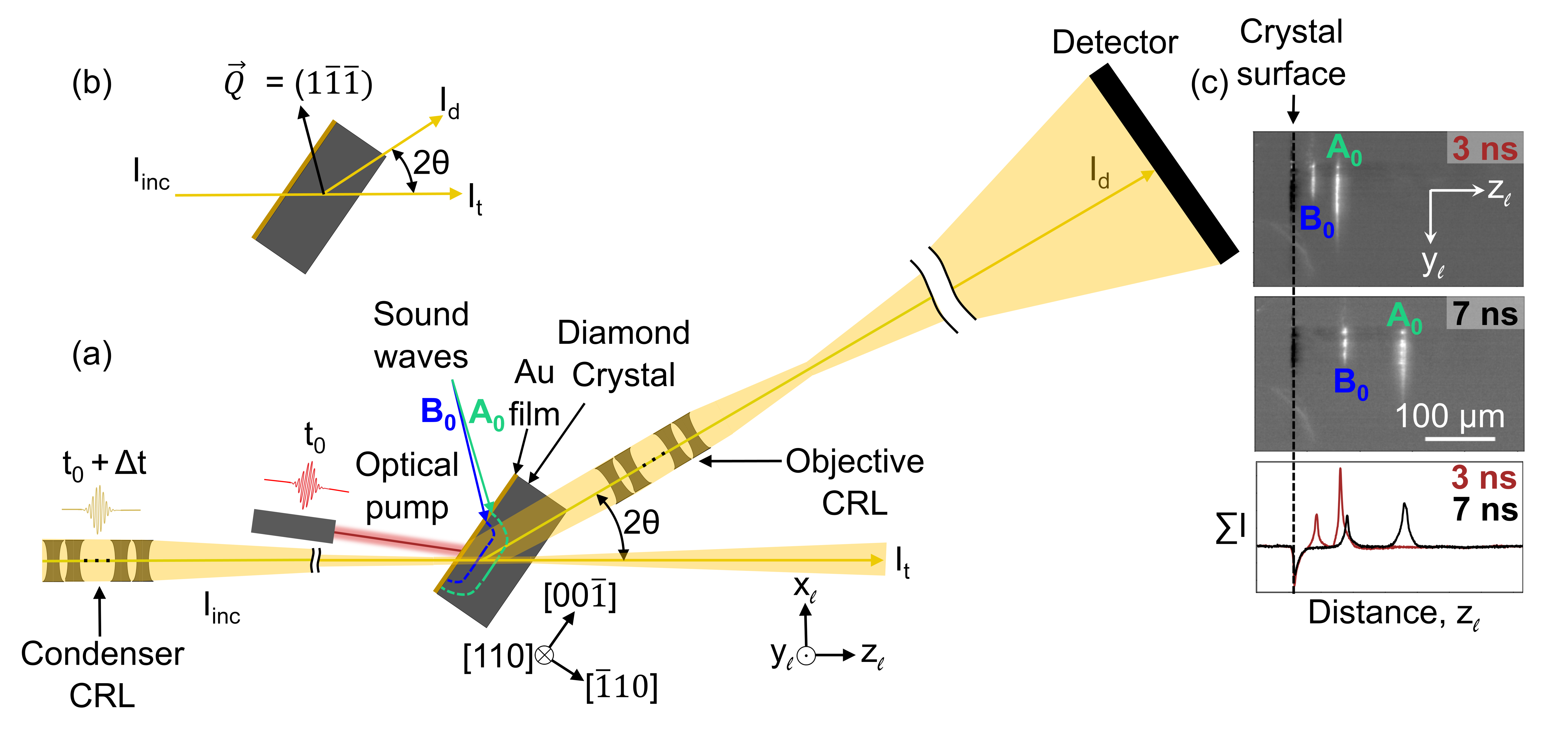}
  \caption{(a) Schematic layout of the DFXM experiment. The incident X-ray beam is condensed in one direction using a compound refractive lens (CRL) to illuminate a layer in the sample, which defines the observation plane (horizontal in this figure). The diamond single crystal is oriented such that diffraction takes place on reflection, $\vec{Q}$.  DFXM images of the observation plane are obtained by using an X-ray objective to magnify the Bragg diffracted X-rays onto a 2D detector. An optical laser pump heats a Au film deposited on the surface of the sample at time $t_0$, leading to thermal expansion and the launching of acoustic waves inside the diamond crystal. The associated local strain variations in the crystalline lattice are imaged at different time-delays $\Delta t$ between the laser-pump and X-ray probe to create a movie of their propagation. Facets of the crystal and a laboratory coordinate system are shown.  (b) Scattering geometry. (c) Experimental DFXM images at $\Delta t$ = 3 ns and 7 ns. Below are graphs of intensity with the vertical $y_{\ell}$-direction integrated out. Two acoustic waves, marked \textbf{A$_0$} (aquamarine) and \textbf{B$_0$} (blue), are seen to propagate towards the right for increasing $\Delta t$. }
  \label{fig:1}
\end{figure}

As illustrated in Fig.~\ref{fig:1}a, we use a fs-duration visible laser to pump the acoustic waves, then use the XFEL pulses to probe the dynamics using an objective lens along the X-ray diffracted beam. By illuminating a thin observation plane in the sample, each image provides a magnified view of the subtle distortions in the crystalline lattice within a 2D slice of the macroscopic mm-sized sample. This allows us to resolve how even subtle deviations from phonons and defects generates local variations (strain and orientation changes) in the lattice. In this work, we acquire movies of the associated dynamics by repeating our experiment over a series of time delays, $\Delta t$, between the optical pump and X-ray probe pulses. 

\begin{figure}[t!] 
\centering
  \centering
  \includegraphics[width=.98\linewidth]{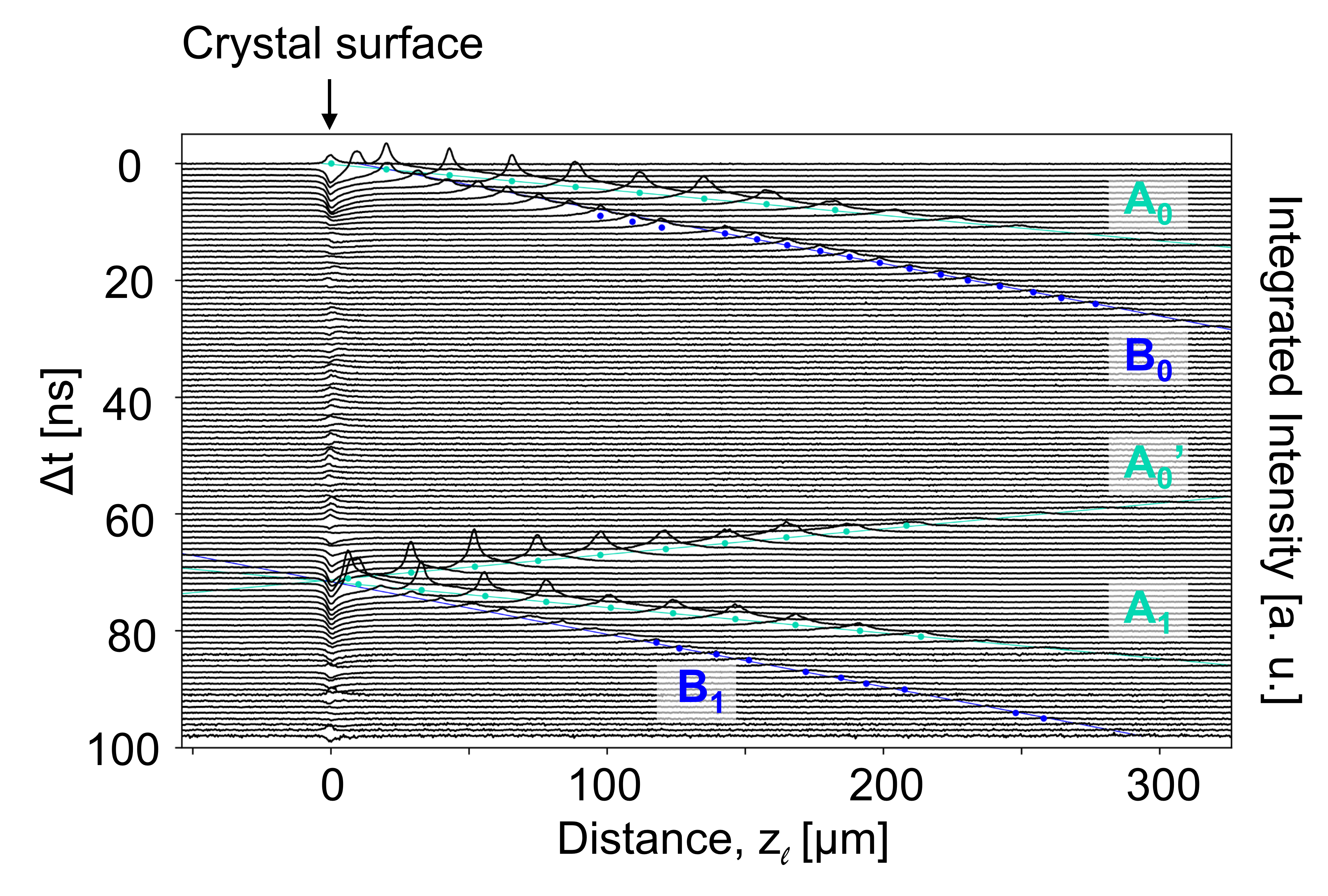}
  \caption{Propagation of the strain waves in diamond over increasing delay times $\Delta t$. For each time step, the 1D trace is generated by integrating the intensity over the vertical $y_{\ell}$-direction, cf. Fig.~\ref{fig:1} (c). The intensity thus maps the temporally weak-beam signal from the strain waves as function of distance along $z_{\ell}$ from the crystal surface, as the wave traverses the observation plane. The plots are vertically offset according to their time-delay.
  The colored dots under each peak signify the positions of each wave's peak, and the straight aquamarine and blue lines plot the linear fits formed by them.
  }
  \label{fig:2}
\end{figure}

Snapshots from a movie (Movie M2 in Suppl. Mater.) spanning time delays of $\Delta t$ = 0 to 100 ns are shown in Fig.  \ref{fig:1}(c). From this movie, it appears that the photo-excitation of the Au film launches two strain waves, labeled $A_0$ and $B_0$, respectively. As predicted (See Fig.~\textbf{\ref{fig:S1}}) the waves are nearly planar over the region we image. To quantify the position of the waves as a function of time, we integrate the image intensities along the $y_{\ell}$-axis and visualize the amplitude and spatial profile as a function of their position along $z_{\ell}$ (see Fig.~\ref{fig:2}). The strain waves A$_0$ and B$_0$ propagate at different velocities as they travel towards the rear face of the diamond crystal (i.e. to the right in Fig.~\ref{fig:2}, beyond the image's field of view). 
The velocity of the fast wave is 18.21 km/s (see Suppl. Mater.), which is consistent with previous reports in the literature of a longitudinal sound velocity of $18.18 \pm 0.03$ km/s in the  $<\!\! 110\!\!>$ directions in diamond \cite{Wang2004}. Assuming the slow wave to propagate in direction $[\bar{1}10]$ as well, the velocity is 8.86 km/s. This is close to a theoretical prediction of 8.95 km/s for the slow transverse acoustic wave along the $<\!\! 110\!\!>$-directions \cite{Krishnan1947}.

\begin{figure}[t!] 
\centering
  \centering
  \includegraphics[width=.98\linewidth]{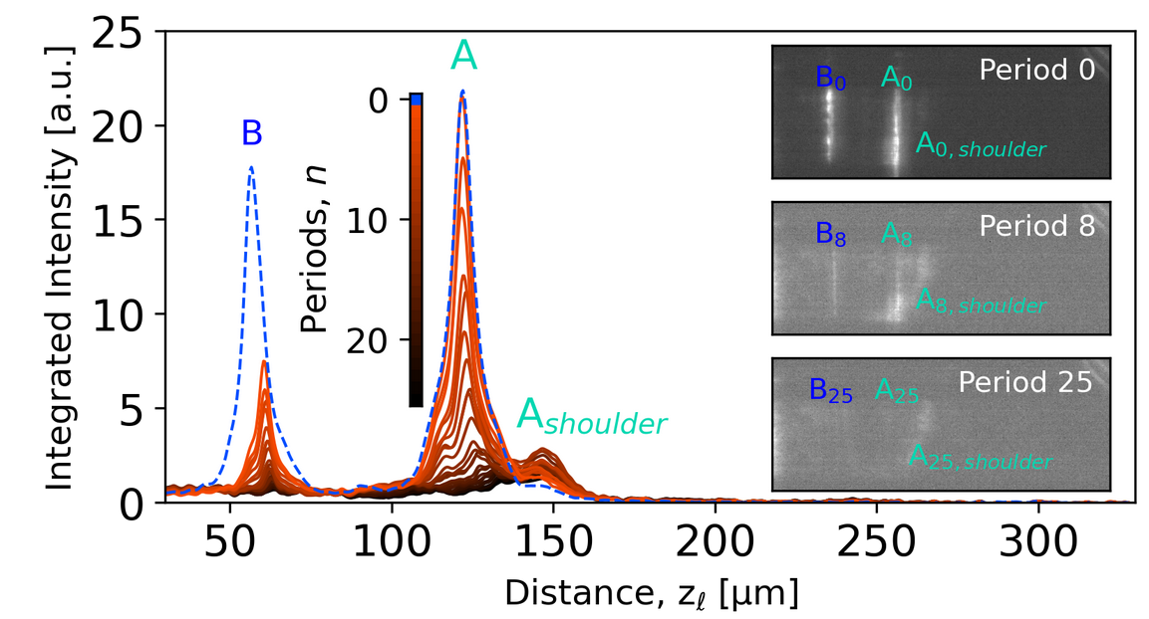}
  \caption{Dispersion of the acoustic waves. Intensity profiles (averaged over $y_{\ell}$) for the fast (A) and slow (B) strains waves at time delays of 5.5 ns + $n \Delta t_p$, where $\Delta t_p$ = 72.47 ns is the time period for the longitudinal wave to return the surface from which it was launched. The curves are generated by integration over a vertical region-of-interest, similar to Fig. \ref{fig:2}. Representative DFXM images are shown on the right for 0th, 8th and 25th period.}
  \label{fig:3}
\end{figure}

Strain wave A$_0$ reflects off the rear surface of the crystal and returns into the field of view (A$_0'$) .
Comparing same positions before and after the component is seen to transmit through the crystal with a nearly soliton structure that scarcely changes the wave's profile as it propagates.
On return to the front Au-coated surface of the crystal, this wave is reflected (A$_1$). 
The multiple ``bounces'' off each surface of the crystal sets up an acoustic cavity that traps the wave between two reflective surfaces as it dissipates energy into the crystal \cite{Lanzillotti2005}.
This is evident from Fig.~\ref{fig:3}, showing snapshots from a movie (Movie M3 in Suppl. Mater.) with time delays of $\Delta t$ = 5.5 ns + $n \Delta t_p$, where $\Delta t_p$ = 72.47 ns is the "period" corresponding to one round-trip. A total of 26 periods were captured.
Figure \ref{fig:3} also shows that that each time wave A is reflected off the Au-coated surface, a new strain wave B is emitted. The apparent formation of a new mode upon each surface reflection suggests that a fixed fraction of energy is transferred from the longitudinal wave to the new transverse waves, with an associated reflectivity of wave A of 89\% per period (see Fig. \textbf{\ref{fig:S3}}). With increasing period number $n$, the intensity profile for strain wave A becomes bimodal.

The contrast in the images in Fig.~\ref{fig:1} (c) was provided by rotating the sample slightly around the $y_\ell$ axis with respect to the Bragg-condition of the bulk crystal. By similar rotation around $z_\ell$ and by varying the scattering angle $2\theta$, three elastic strain components can be mapped by DFXM with a strain resolution better than $10^{-4}$ \cite{Poulsen2021}.  Determining the direction of displacement of the waves in this way corroborates that A is a longitudinal wave (see Suppl. Mater.).  Moreover, we find that the spatial variation of the strain-wave contrast is well described by a combined thermomechanical and X-ray  geometrical optics forward simulation, see Fig. \ref{fig:S8}.

This study was performed with an \emph{ad hoc} experimental setup. With available X-ray optics the spatial resolution in DFXM at the synchrotron is 100 nm \cite{Poulsen2018}. By translating the sample perpendicularly to the X-ray beam and by varying the sample orientation movies can be made with a high resolution in both 3D direct space and 3D reciprocal space, enabling \emph{e.g.} a full spatiotemporal mapping of the phonon spectrum (see Suppl. Mater.). 

Until now, ultrafast work on phonon dynamics has primarily been \emph{spectroscopic}, using time-domain Brillouin and Raman scattering \cite{Bojahr2013}, or electron and X-ray \emph{diffraction} studies \cite{Lindenberg2000, Larsson2002,  Lemke2018}.
Spatially resolved studies have been restricted to thin foils using ultra-fast electron microscopy \cite{Lobatsov2005,Cremons2016}, to surface acoustic waves \cite{Shilo2002}, or to optically transparent samples \cite{Ofori-Okai2014}. 
In contrast, the method presented above can image traveling acoustic waves, generated by packets of phonon modes, in the bulk of optically opaque materials, with a high and symmetry-selective strain sensitivity. It is well suited to probe the interaction between strain waves and structural elements such as dislocations, grain boundaries or voids. This is of critical importance to understand phonon lensing in meta-materials and photonic crystals \cite{Cummer2016}, and for thermal engineering studies that seek to control the flow of heat in thermoelectric materials \cite{Chen2003}. We also see opportunities for this approach in geosciences, extending it to polycrystalline materials to test seismological models of sound propagation in planetary materials \cite{Sha2020}. 

Exploiting that acoustic strain wave generation is   a reversible process, the results represent averages over several images taken at the same delay time.   However, the intensities in the individual raw images before averaging are $\sim$200 counts/pixel (see Movie M1), with a SNR of $\sim$8.5. Moreover, the X-ray scattering signal from nearly all materials is larger than that of diamond. Hence, each pump event may be followed by not one but a train of hundreds of X-ray probe pulses - as provided by some XFELs.  Uniquely, this opens opportunities for for visualising stochastic and \textit{irreversible} structural processes in real-time on the sub-µs timescale. Such processes are ubiquitous in materials science, \emph{e.g.} martensitic phase transformations in steel, domain switching in ferroelectrics, and dielectric break-down.

\bibliography{main_science}{}

\begin{thebibliography}{10}

\bibitem{Cahn1991}
R.~Cahn, P.~Haasen, E.~Kramer, {\it Materials Science and Technology\/} (VCH,
  Weinheim, 1991).

\bibitem{Lauridsen2001}
E.~M. Lauridsen, S.~Schmidt, R.~M. Suter, H.~F. Poulsen, {\it J. Appl.
  Crystallogr.\/} {\bf 34}, 744 (2001).

\bibitem{Suter2006}
R.~M. Suter, D.~Hennessy, C.~Xiao, U.~Lienert, {\it Rev. Sci. Instrum.\/} {\bf
  77}, 123905 (2006).

\bibitem{Johnson2008}
G.~Johnson, A.~King, M.~G. Honnicke, J.~Marrow, W.~Ludwig, {\it J. Appl.
  Crystallogr.\/} {\bf 41}, 310 (2008).

\bibitem{Larson2002}
B.~C. Larson, W.~Yang, G.~E. Ice, J.~D. Budai, J.~Z. Tischler, {\it Nature\/}
  {\bf 415}, 887 (2002).

\bibitem{Ice2011}
G.~E. Ice, J.~D. Budai, J.~W.~L. Pang, {\it Science\/} {\bf 334}, 1234 (2011).

\bibitem{Simons2015}
H.~Simons, {\it et~al.\/}, {\it Nat. Commun.\/} {\bf 6}, 6098 (2015).

\bibitem{Poulsen2018}
H.~F. Poulsen, {\it et~al.\/}, {\it J. Appl. Crystallogr.\/} {\bf 51}, 1428
  (2018).

\bibitem{Schmidt2004}
S.~Schmidt, {\it et~al.\/}, {\it Science\/} {\bf 305}, 229 (2004).

\bibitem{Zhang2020}
J.~Zhang, {\it et~al.\/}, {\it Acta Mater.\/} {\bf 191}, 211 (2020).

\bibitem{Margulies2001}
L.~Margulies, G.~Winther, H.~Poulsen, {\it Science\/} {\bf 291}, 2392 (2001).

\bibitem{King2008}
A.~King, G.~Johnson, D.~Engelberg, W.~Ludwig, J.~Marrow, {\it Science\/} {\bf
  321}, 382 (2008).

\bibitem{Simons2018}
H.~Simons, {\it et~al.\/}, {\it Nat. Mater.\/} {\bf 17}, 814 (2018).

\bibitem{Dresselhaus-Marais2021}
L.~E. Dresselhaus-Marais, {\it et~al.\/}, {\it Sci. Adv.\/} {\bf 7}, eabe8311
  (2021).

\bibitem{Langer2016}
M.~Langer, F.~Peyrin, {\it Osteoporos. Int.\/} {\bf 27}, 441–455 (2016).

\bibitem{Holstad2022}
T.~S. Holstad, {\it et~al.\/}, {\it J. Appl. Crystallogr.\/} {\bf 55}, 112
  (2022).

\bibitem{Wang2004}
S.~F. Wang, Y.~F. Hsu, J.~C. Pu, J.~C. Sung, L.~G. Hwa, {\it Mater. Chem.
  Phys.\/} {\bf 85}, 432 (2004).

\bibitem{Krishnan1947}
R.~S. Krishnan, {\it Proc. Natl. Acad. Sci. India A\/} {\bf 26}, 399 (1947).

\bibitem{Lanzillotti2005}
N.~D.~L. Kimura, A.~Fainstein, B.~Jusserand, {\it Phys. Rev. B\/} {\bf 71},
  041305(R) (2005).

\bibitem{Poulsen2021}
H.~F. Poulsen, L.~E. Dresselhaus-Marais, M.~A. Carlsen, C.~Detlefs, G.~Winther,
  {\it J. Appl. Crystallogr.\/} {\bf 54}, 1555 (2021).

\bibitem{Bojahr2013}
A.~Bojahr, {\it et~al.\/}, {\it Opt. Express\/} {\bf 21}, 21188 (2013).

\bibitem{Lindenberg2000}
A.~M. Lindenberg, {\it et~al.\/}, {\it Phys. Rev. Lett.\/} {\bf 84}, 111
  (2000).

\bibitem{Larsson2002}
J.~Larsson, {\it et~al.\/}, {\it Appl. Phys. A\/} {\bf 75}, 467 (2002).

\bibitem{Lemke2018}
H.~T. Lemke, {\it et~al.\/}, {\it ACS Omega\/} {\bf 3}, 9929 (2018).

\bibitem{Lobatsov2005}
V.~A. Lobastov, R.~Srinivasan, A.~H. Zewail, {\it Proc. Natl. Acad. Sci.
  U.S.A.\/} {\bf 102}, 7069 (2005).

\bibitem{Cremons2016}
D.~R. Cremons, D.~A. Plemmons, D.~J. Flannigan, {\it Nat. Commun.\/} {\bf 7},
  11230 (2016).

\bibitem{Shilo2002}
D.~Shilo, E.~Lakin, E.~Zolotoyabko, J.~Härtwig, J.~Baruchel, {\it Synchrotron
  Radiat. News\/} {\bf 15}, 21 (2002).

\bibitem{Ofori-Okai2014}
B.~K. Ofori-Okai, P.~Sivarajah, S.~M. Teo, C.~A. Werley, K.~A. Nelson, {\it
  Ultrafast Nonlinear Imaging and Spectroscopy II\/} (2014), vol. 9198, p.
  919813.

\bibitem{Cummer2016}
S.~A. Cummer, J.~Christensen, A.~Alù, {\it Nat. Rev. Mater.\/} {\bf 1}, 16001
  (2016).

\bibitem{Chen2003}
G.~Chen, M.~Dresselhaus, G.~Dresselhaus, J.-P. Fleurial, T.~Caillat, {\it Int.
  Mater. Rev.\/} {\bf 48}, 45 (2003).

\bibitem{Sha2020}
G.~Sha, M.~Huang, M.~J.~S. Lowe, S.~I. Rokhlin, {\it J. Acoust. Soc. Am.\/}
  {\bf 147}, 2442 (2020).

\bibitem{Dresselhaus2022}
L.~E. Dresselhaus-Marais, {\it et~al.\/}, {\it arXiv:2210.08366v2\/} {\bf
  [cond-mat.mtrl-sci]} (2022).

\end{thebibliography}
\bibliographystyle{Science}

\section*{Acknowledgements}

Financial support was provided by the Villum FONDEN (grant no. 00028346) and the ESS lighthouse on hard materials in 3D, SOLID, funded by the Danish Agency for Science and Higher Education (grant number 8144-00002B). Moreover, H.F.P. and H.S. acknowledges support from the European Research Council (Advanced grant no 885022 and Starting grant no 804665, respectively). We further acknowledge that this work was performed in part under the auspices of the US Department of Energy by Lawrence Livermore National Laboratory under contract DE-AC52-07NA27344. Initial contributions from LEDM were also funded by the support of the Lawrence Fellowship at LLNL. TMR acknowledges funding from the European Union’s Horizon 2020 research and innovation programme under the Marie Skłodowska-Curie grant agreement No 899987.

\section*{Additional information}

\section*{Materials and Methods}
The X-ray microscope established at the X-ray Correlation Spectroscopy (XCS) instrument at the LCLS XFEL during beamtime in June 2021 was similar to existing DFXM instruments at synchrotrons \cite{Simons2015,Poulsen2018}. The full details of the optical and hardware set-up are presented in Ref \cite{Dresselhaus2022}. In brief, the 10.1 keV X-ray pulse was monochromatised to provide an energy bandwidth of $\Delta E/E = 10^{-4}$.  The incident beam was focused horizontally to a thickness of 3.9 µm by a compound refractive lens (CRL) 
condenser placed 3.43 m before the sample.
The objective CRL comprised 33 Be lenslets with a 50 µm radius, and a resulting focal length of 0.205 m and a numerical aperture (FWHM) of $ 8.5 \times 10^{-4}$\cite{Poulsen2018}. The sample-detector distance was 6.83 m. The 2D detector comprised an Andor Zyla 5.5 sCMOS camera with 6.5 µm pixelsize coupled to a scintillator screen by visual optics. 
The resulting magnification of the diffraction imaging system was determined to be $\sim$30. 

 Here, we combine DFXM with a pump-probe scheme. A 300 nm gold film was sputter coated on the polished $(\bar{1}10)$-facet of an Element 6 diamond single crystal grown via  chemical vapour deposition and laser cut to dimensions of 1x2x0.66 mm$^3$ 
 (see Fig.~\ref{fig:1}). A 15 nm Ti film was used as an adhesion layer. An optical laser pump from a Ti:Sapphire laser system with a wavelength of 800 nm, an energy per pulse of 100 µJ and a pulse duration of 50 fs  was used to induce ultrafast heating and expansion within a spot with a diameter of 150 µm (FWHM) in the Au layer, thereby launching strain waves into the diamond. An X-ray probe with an energy per pulse of 1.6 mJ and a pulse duration of 50 fs, was used to image the propagation of the strain waves. For the $(1 \bar{1} \bar{1})$ Bragg-reflection in diamond, the scattering angle is 2$\theta$ = 35.04$^{\circ}$ for 10.1 keV X-rays. In this setting the effective pixel size in the illuminated plane is 215.8 nm $\times$ 375.9 nm along $y_{\ell}$ and $z_{\ell}$, respectively.  A region-of-interest was set corresponding to a field of view of the microscope within the illuminated layer  of 221 µm $\times$ 385 µm. The FWHM thickness (along $x_{\ell}$; see Fig.~\ref{fig:1}) of the observation plane formed by the incident X-rays was $\Delta x$ $\approx$ 3.9 µm. The spatial resolution in the two orthogonal directions is much better, essentially given by the detector pixel size.  For details of the spatial and angular resolution see Suppl. Mater.  Weak-beam contrast was used with an angular off-set of +0.0764 mrad in $\phi$ with respect to the orientation where the strain-free parts of the diamond crystal is in the Bragg condition.

 Images were acquired with the X-ray probe beam having time delays of $\Delta t$ with respect to  the arrival time of the optical pump. The signal-to-noise ratio in individual images is $\sim$8.5. To improve signal-to-noise, we collected 240 frames at each time delay. By default the images in this work (and subsequent datasets) represent the average of the 240 frames per time-step, corrected for background. An example of such a set of frames is provided as Movie M1. No beam damage was observed while exposing a fixed volume for three hours with the LCLS repetition rate of 120 Hz.  
 
 In Figs.~\ref{fig:2} and \textbf{\ref{fig:S2}}, there is a jump in the position of strain-wave B between 11 and 12 ns. This was due to an anomalously large time-step of the timing-tool at LCLS, and not a real effect.

\newpage

\section*{Supplementary materials}
\renewcommand{\figurename}{\textbf{Figure}}
\renewcommand{\tablename}{\textbf{Table}}
\setcounter{figure}{0}
\renewcommand{\thesection}{S\arabic{section}}
\renewcommand{\thetable}{\textbf{S}\textbf{\arabic{table}}}
\renewcommand{\thefigure}{\textbf{S}\textbf{\arabic{figure}}}
\newcounter{SIfig}
\renewcommand{\theSIfig}{S\arabic{SIfig}}
\baselineskip12pt

Supplementary Text including Figs. \textbf{S1} to \textbf{S11}.\\
Movies M1 to M3.\\

\subsection*{Real-time X-ray microscopy movies of the acoustic waves}

Movie M1:  Gallery of 120 randomly selected raw images for a time delay of 77.97 ns. This illustrates the intensity fluctuations between individual X-ray pulses.

Movie M2: DFXM movie of the structural changes in observation plane during the first 100 ns after the ultrafast heating of the Au foil. 

Movie M3: DFXM movie of the structural changes in observation plane during the first 1800 ns after the ultrafast heating of the Au foil. Shown are snapshots acquired at times 5.5 ns + $n\Delta t_p$, where $n$ is an integer and $\Delta t_p$ = 72.47 ns is the period corresponding to the fast strain wave travelling from the Au coated surface to the free surface and back.

\subsection*{Geometry of the experiment}

\begin{figure}[!ht]
\centering
  \centering
  \includegraphics[width=.98\linewidth]{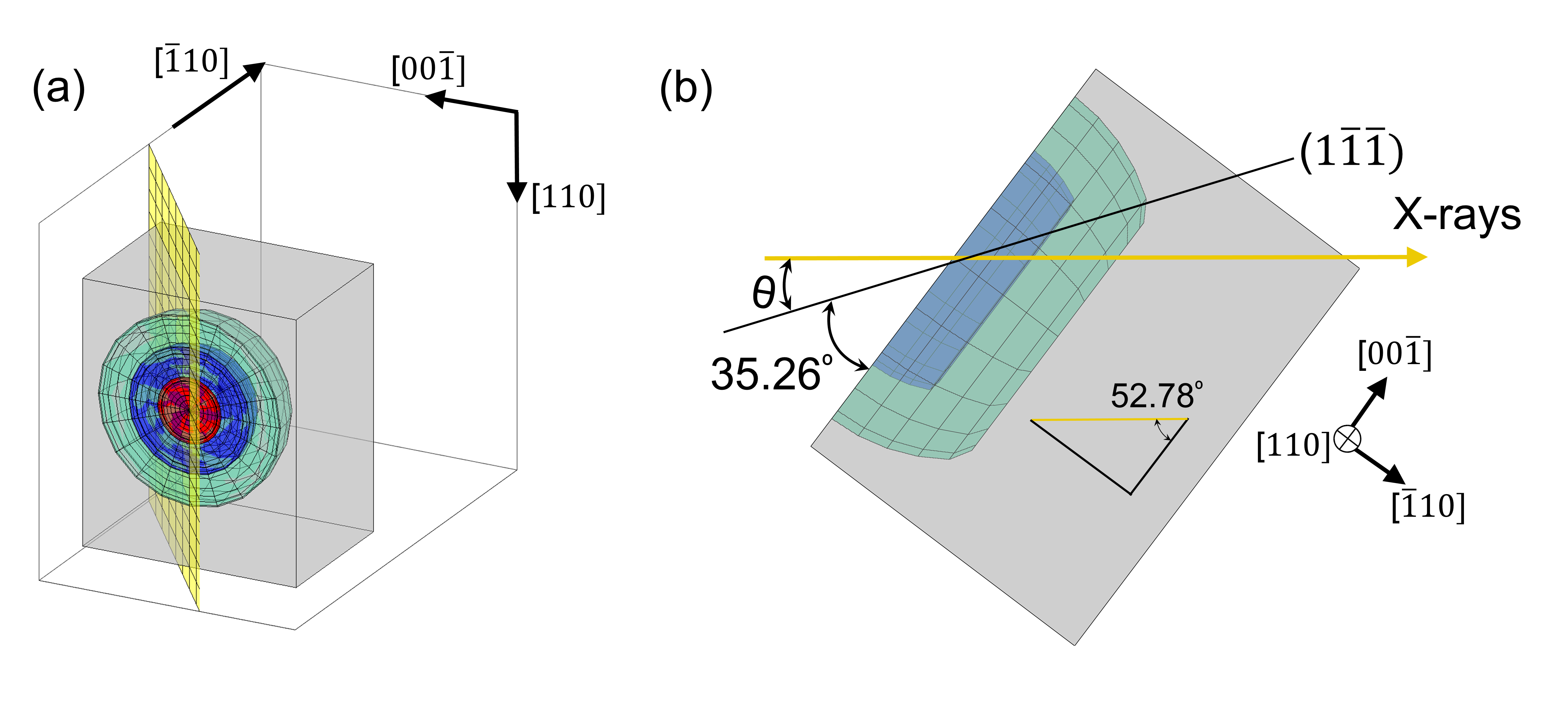}
  \caption{3D Schematic of the geometry of the experiment. The coordinate systems are the same as in Fig.~\ref{fig:1}. (a) The transparent blue and aquamarine surfaces represent the transverse and longitudinal acoustic strain waves, respectively. The red area represents the 150 µm optical laser spot on the Au-coated surface.  Within a region given by the optical laser spot size the strain waves travel as planar waves. (b) Top-down view on the the diamond crystal, which is oriented such that Bragg-scattering occurs from the $(1\bar{1}\bar{1})$-planes. Hence, the angle between the observation plane (the layer illuminated by the incident X-ray beam; yellow transparent surface in this figure) and the $(1\bar{1}\bar{1})$-planes is $\theta$ = 17.52$^{\circ}$. The angle between $(1\bar{1}\bar{1})$-planes and the $(\bar{1}10)$-planes is 35.26$^{\circ}$. Thus, the relation between the observed distance in the observation plane $z_{\ell}$ and the actual distance $z_{\textrm{sw}}$ the waves travel along $[\bar{1}10]$ is given by $z_{\textrm{sw}} = z_{\ell} \cdot \sin(52.78^{\circ}) $.}

\refstepcounter{SIfig}\label{fig:S1}
\end{figure}

The geometry of the experiment is sketched in Fig.\ref{fig:1}. The X-ray beam illuminates a 2D-sheet (the observation plane) of the diamond crystal, while the strain waves have an anisotropic three-dimensional structure. In Fig.~\textbf{\ref{fig:S1}} we provide a 3D illustration of the two strain waves, for a given time delay, assuming the waves travel in a homogeneous media. The figure illustrates that the waves are near planar within the central part of the field of view - consistent with the DFXM images. 

\begin{figure}[!ht] 
\centering
  \centering
  \includegraphics[width=1\linewidth]{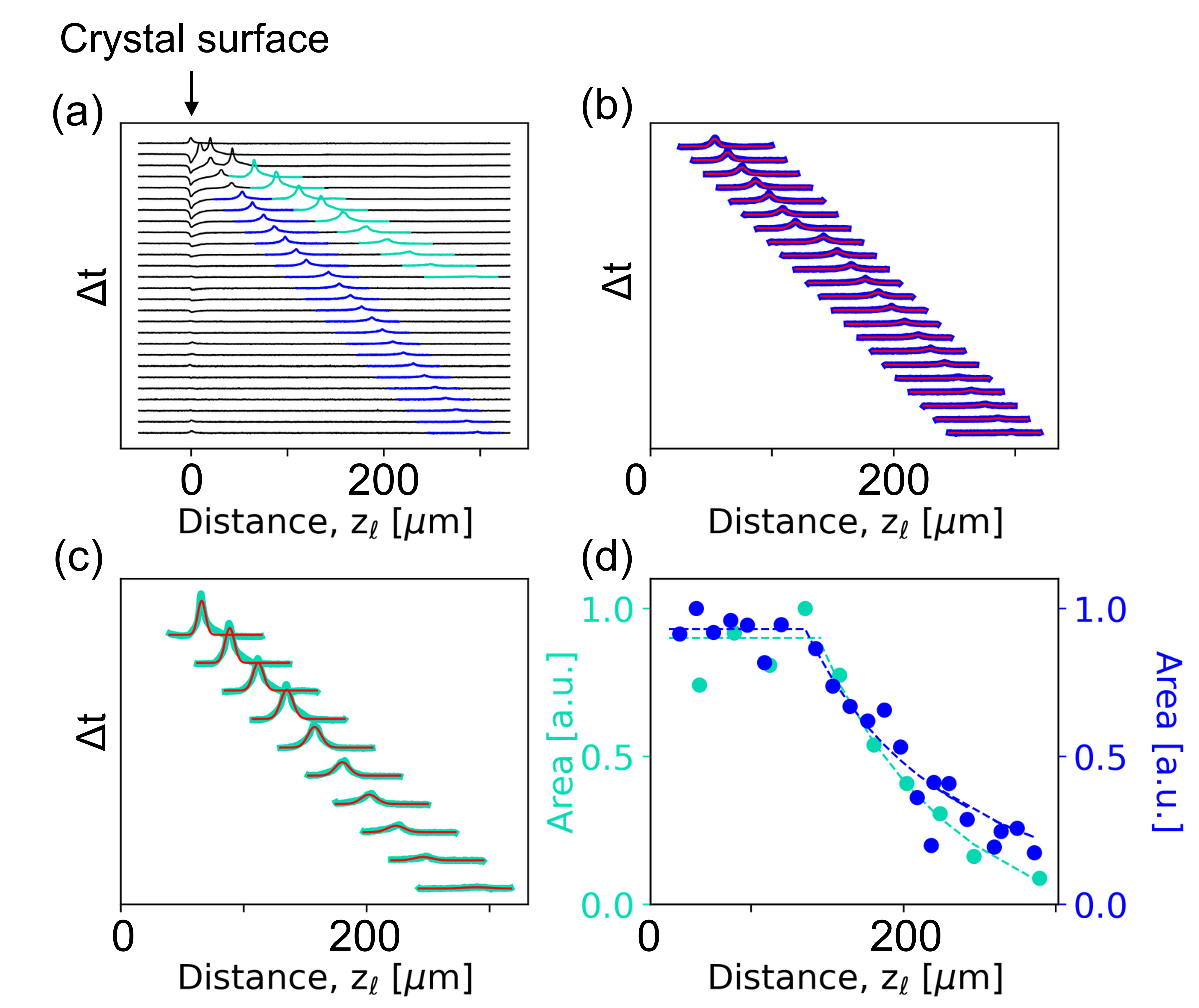}
  \caption{(a) Intensity curves (averaged over $y_{\ell}$) as function of distance $z_{\ell}$ (same data as in Fig.~\ref{fig:2}). (b,c) Gaussian fits (red curves) to the experimental data for strain waves B and A, respectively. (d) The areas of the Gaussian fits to peak B and peak A are plotted as function of distance (dots). In both cases data are normalised to 1. The dashed lines are best fits to models, where areas are constant in the initial 150 µm, and then  decreases as $z_{\ell}^{-2}$ for larger z$_{\ell}$. }

\refstepcounter{SIfig}\label{fig:S2}
\end{figure}

Figure \textbf{\ref{fig:S2}} displays the evolution of the intensity profiles (averaged over $y_{\ell}$) of the two strain waves in Fig.~\ref{fig:2} during the initial 30 ns. The intensity profiles  are seen to be relatively constant over the initial 150 µm, corresponding to the diameter of the laser spot on the surface. This is the region in Fig.~\textbf{\ref{fig:S1}} where the strain wave is planar. After 150 µm, the strain wave intensity decreases as $z_{\ell}^{-2}$, consistent with the energy being distributed over a spherical surface.

The sound velocity for strain wave A can be determined with high accuracy by the time $\Delta t$ it takes to perform 26 periods and the sample thickness. This result was used to calibrate the magnification and subsequently to determine the velocity of strain wave B.

\newpage
\FloatBarrier
\subsection*{Transfer of energy from the longitudinal to the transverse wave}

We observe that for all periods $n$, the strain wave A is reflected upon the Au-deposited surface, and generating a new slow strain wave B$_n$ in the process (the same process may occur on the opposite surface, but this was outside of the field of view, and thus not directly observed). As the intensity profiles of the B$_n$ waves are identical as function of time delay since their creation, we infer that they are all transverse waves of the same type as B$_1$. 

\begin{figure}[!ht] 
\centering
  \centering
  \includegraphics[width=.98\linewidth]{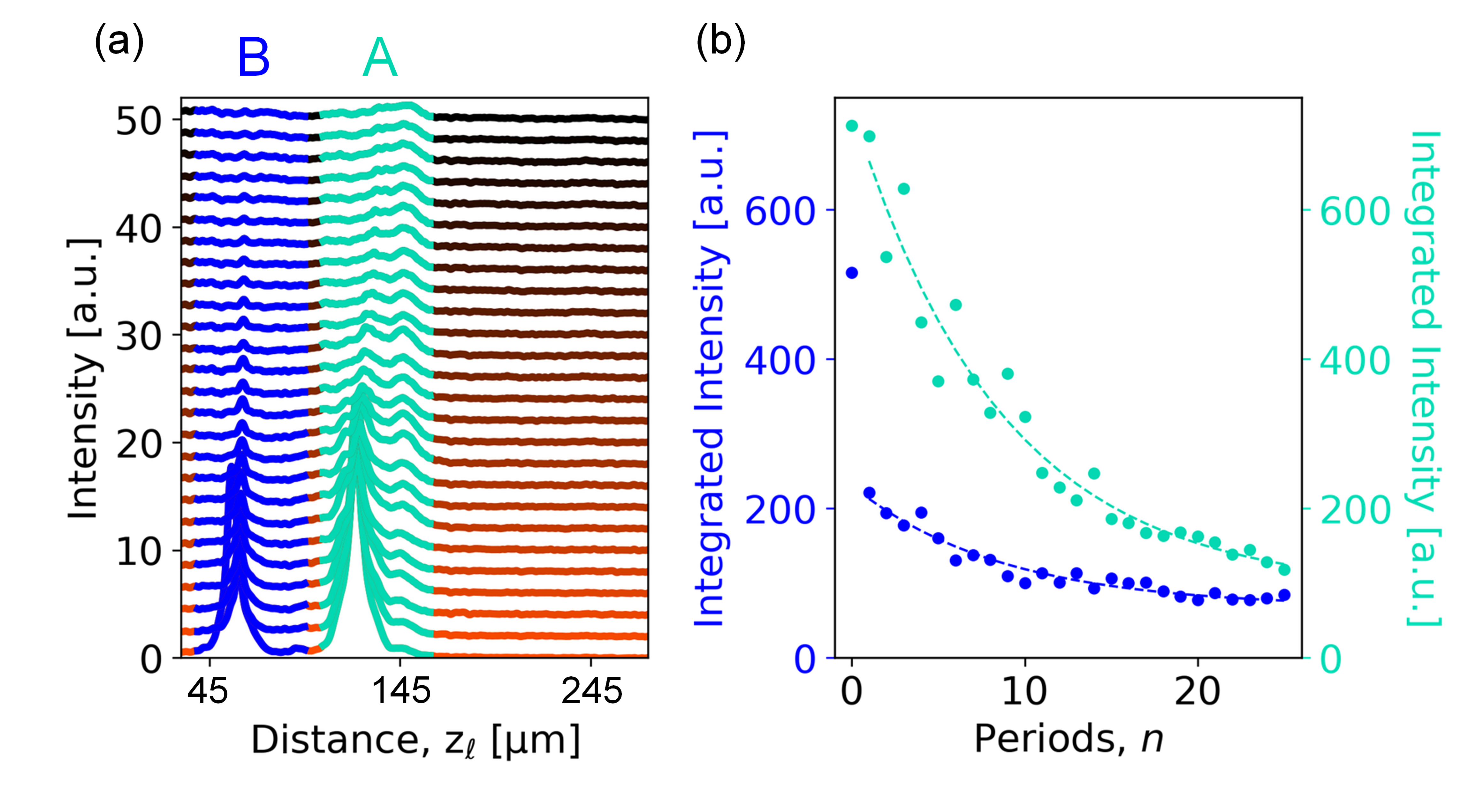}
  \caption{(a) Intensity profiles for 26 equi-distant time delays of $\Delta t$ = 5.5 ns +$ n\Delta t_p$, where $n$ is an integer and  $\Delta t_p= 72.47$ ns is the time period it takes for the longitudinal strain wave A to travel from the Au-coated surface to the free surface and back. The curves are offset by ~$2n$*a.u. for ease of visualisation. The curves are integrated within the regions $z_{\ell} \in [39 \; 95]$ µm and  $z_{\ell} \in [105 \; 161]$ µm for peak B and A, respectively (indicated by color). (b) The resulting integrated intensities as a function of number of periods, $n$ (dots). The two dashed lines represent a simultaneous fit to functions $I_i = a_i e^{-bn} + c_i, i =$ A, B with $c_i$ being background terms. The quality of this fit corroborates a hypothesis that both peaks exhibit exponential decays with the same constant $b=0.116$.}

\refstepcounter{SIfig}\label{fig:S3}
\end{figure}

We make the hypothesis that the creation of transverse waves in each period is associated with a transfer of a fixed fraction of energy from the longitudinal wave to the new transverse waves. As illustrated in Fig.~\textbf{\ref{fig:S3}} a combined fit to the decay of the longitudinal wave and the decaying relative intensities of the new transversal waves corroborates this hypothesis. The fitted values are provided in Table~\ref{table:s1}. The fitted decay rate $b=0.116$ corresponds to a constant transfer of energy from the longitudinal to the transverse waves and a "reflectivity" for each period of $R = e^{-b} = 0.89$.

\begin{table}
\caption{The optimised parameters $a_i$, $b$ and $c_i$, based on the fit illustrated in Fig.~\textbf{\ref{fig:S3}}.}
\label{table:s1}
\begin{center}
\begin{tabular}{||c |c |c||} 
 \hline
  $i$ & A & B  \\ [0.5ex] 
 \hline\hline
  $a_i$ & 646.2 & 163.2  \\ 
 \hline
  $b$ & 0.116 & 0.116 \\
 \hline
  $c_i$ & 89.8 & 67.5  \\ [1ex] 
 \hline
\end{tabular}
\end{center}
\end{table}

\FloatBarrier
\subsection*{Mapping strain components}

DFXM is sensitive to the individual components of the displacement gradient tensor field $\mathbf{F}(\vec{r}, t)$ \cite{Poulsen2021}, which can be expressed in terms of an orientation field $\mathbf{\Omega}(\vec{r},t)$, characterising grains and domains, and an elastic strain tensor field, $\boldsymbol{\epsilon}(\vec{r}, t)$, related to the local stresses by Hooke's law. 

With the set-up illustrated in Fig.~\ref{fig:1}, where we diffract from the $ \vec{Q} =(1 \bar{1} \bar{1})$ lattice planes, contrast can be provided in three ways (see Fig. 1):
\begin{itemize}
    \item \emph{Rotation around $y_{\ell}$} by angle $\phi$. Known as a rocking scan, this probes a shear strain: the displacement of $\vec{Q}$ along the $[ \bar{1}1 \bar{2}]$ direction. (For a longitudinal wave along $[\bar{1}10]$ the component of the strain $\epsilon_L$ that displaces $\vec{Q}$ along direction $[ \bar{1}1 \bar{2}]$ is $\epsilon_L$ cos(54.74$^{\circ}$), where 54.74$^{\circ}$ is the angle between $[\bar{1}10]$ and $[1 \bar{1} 2]$. This is the contrast mode used in Figs.~\ref{fig:1}-\ref{fig:3}, \textbf{\ref{fig:S2}}-\textbf{\ref{fig:S3}}, and \ref{fig:S7}-\ref{fig:S10}).
    \item \emph{Rotation around $z_{\ell}$} by angle $\chi$. Known as a rolling scan, this probes another shear strain: the displacement of $\vec{Q}$ along the $[1 1 0]$ direction.
    \item \emph{Variation of axial strain}. A combined $2\theta-\phi$ scan probes the axial strain (the elongation) in direction $\vec{Q}$. 
\end{itemize}

A longitudinal wave travelling in direction $[\bar{1} 1 0]$ will exhibit strain in the same direction. If the instrumental blurring is negligible this is visible in both $\phi$ and axial strain scans, while $\chi$-scans are not sensitive to this strain component. The same is true for a transverse wave travelling in direction $[\bar{1} 1 0]$ with a strain in direction $[0 0\bar{1}]$. On the other hand a  transverse wave travelling in direction $[\bar{1} 1 0]$ with a strain in direction $[1 1 0]$ provides only contrast when $\chi$ is offset from 0.
With a more quantitative description \cite{Poulsen2021} it appears that DFXM can identify both the direction of propagation and the direction of the displacement of the acoustic waves. 

Due to mechanical constrains in the \emph{ad hoc} setup, $\chi$ and $2\theta$ could not be varied.  What can be deduced is that the observation by $\phi$-contrast of the two strain waves is consistent with a longitudinal wave and a slow transverse wave (with a displacement in direction  [00$\bar{1}$]) as suggested by the speed of sound values. From this follows that a third fast transverse wave (with displacement in direction [110]) might be created, but is invisible with the configuration used. \newline

\subsection*{Comparing experimental and simulated strain-wave profiles}

\begin{figure}[!ht] 
\centering
  \centering
  \includegraphics[width=1\linewidth]{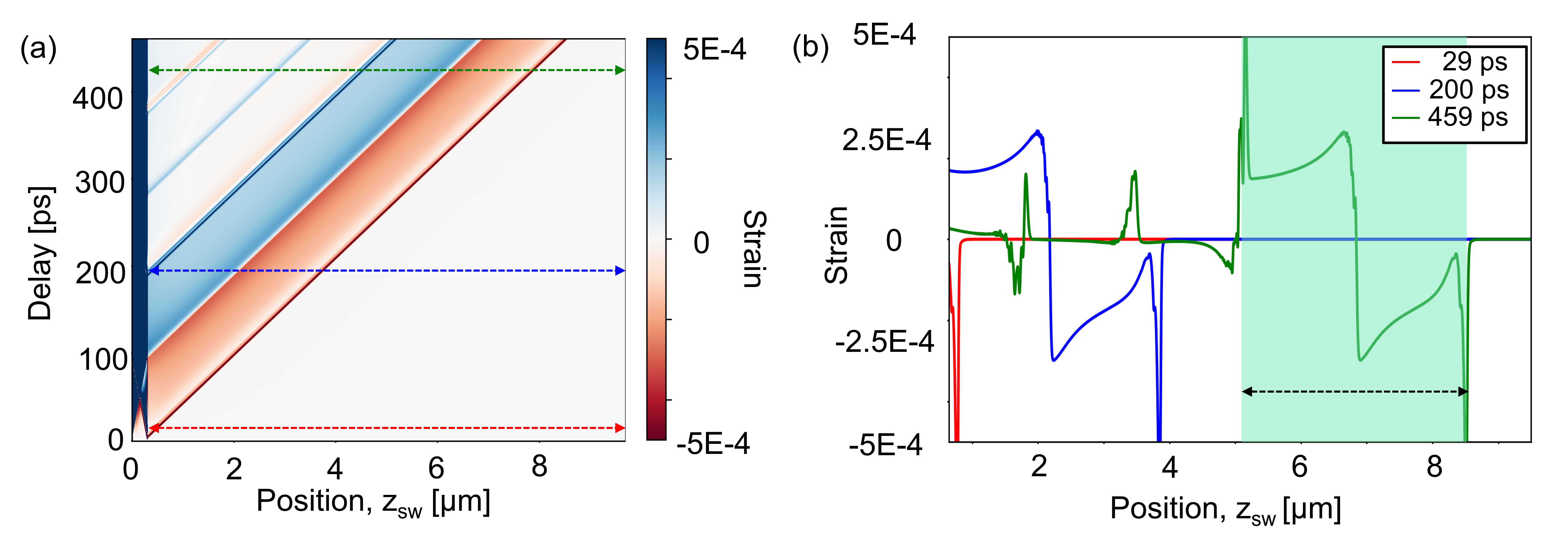}
  \caption{Strain-wave profile in a diamond single crystal as a function of time delay
from laser pulse heating, as computed using a 1D thermomechanical  model. (a) 2D map of strain versus depth and time delay. (b) 1D plots of the strain profile in diamond at different time delays (indicated with dashed lines in (a)). The spatial extent of the part of the strain wave that is visible in DFXM at 459 ps (indicated in transparent green) is about 3.5 µm.}
  \label{fig:S4}
\end{figure}

In this section we will compare experimental DFXM images to geometrical-optics-based forward modelling projections. The forward model takes a thermomechanical model of the strain-wave as its input. The  thermomechanical model for ultrafast dynamics applied, udkm1Dsim (D. Schick. \emph{Comput. Phys. Commun.} \textbf{266}, 108031 (2021)),
is one-dimensional and represents a longitudinal wave.  The strain profile arising for a 300 nm Au film on 15 nm Ti on diamond, excited by a 100 $\mu$J optical laser pulse with a 150 $\mu$m FWHM diameter spotsize on the sample surface, is presented in Fig.~\ref{fig:S4}. Once the strain wave is formed its profile does not change with time in this model (no dispersion).

\begin{figure}[!ht] 
\centering
  \includegraphics[width=0.8\linewidth]{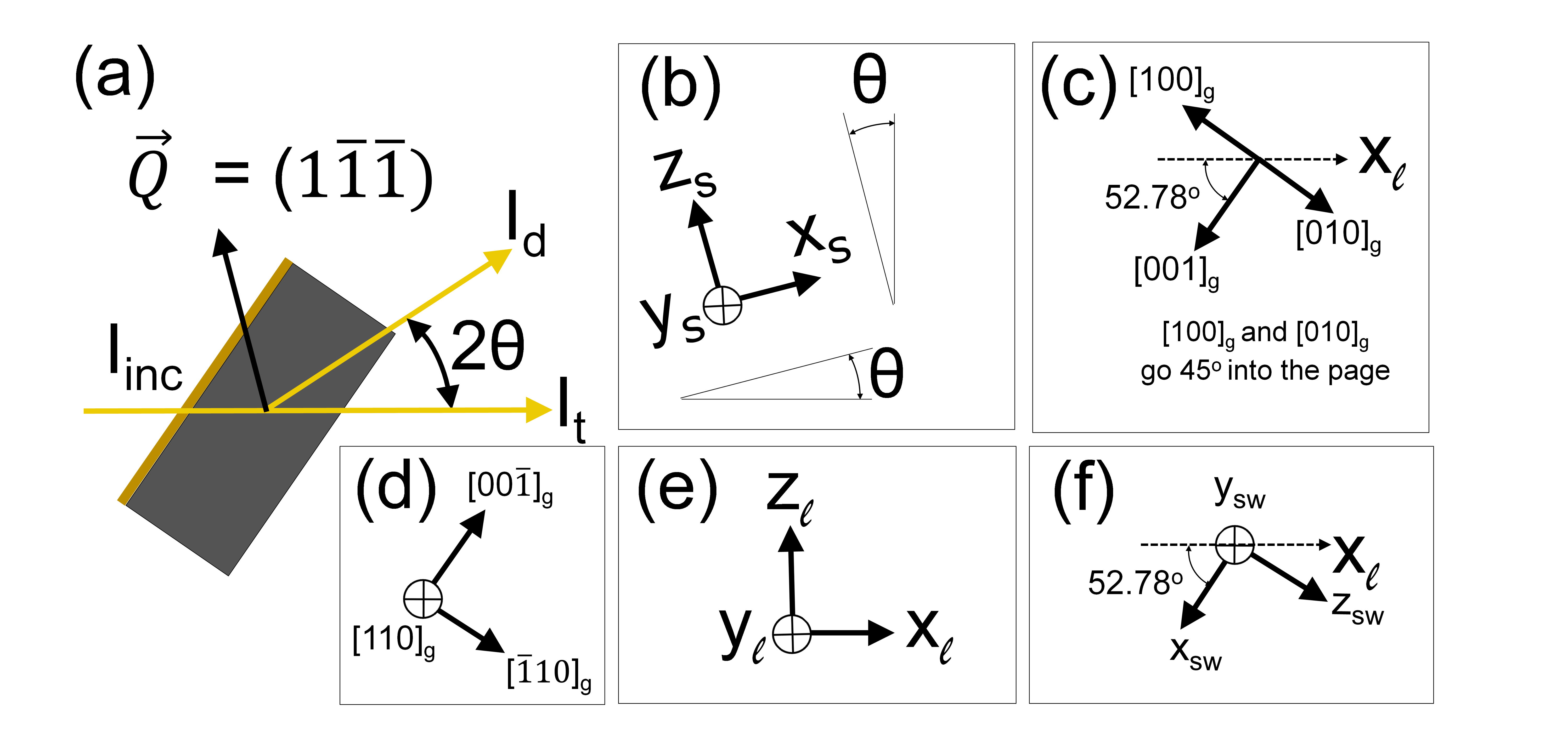}
  \caption{Coordinate systems used, with reference the conventions used in Ref.~\cite{Poulsen2021}. (a) Experimental geometry (horizontal plane) with $\vec{Q}$ the diffraction vector and $I_{\textrm{inc}}$, $I_{\textrm{d}}$ and $I_{\textrm{t}}$ being the incoming, diffracted and transmitted beams, respectively. (e) The laboratory coordinate in the synchrotron convention used in Ref.~\cite{Poulsen2021}. (In the rest of the paper, outside this section, with direct reference to the geometrical optics model, we use the XFEL convention.) (b) The associated sample coordinate system rotated $\theta$ around the $y_{\ell}$-axis relative to the laboratory coordinate system. (c) The grain coordinate system projected down on the plane of the figure (the x- and y-axes go 45$^{\circ}$ into the page). (d) The grain coordinate system, with directions parallel to the diamond crystal's facets.  (f) An additional strain-wave coordinate system, with $x_{\textrm{sw}}$ along $[001]_{\textrm{g}}$ and $z_{\textrm{sw}}$ along $[\bar{1}10]_{\textrm{g}}$.}
  \label{fig:S5}
\end{figure}

The DFXM forward projection model used is a simple adaptation of the geometrical optics code presented in Ref.~\cite{Poulsen2021}. This uses a synchrotron convention for the laboratory coordinate system~\cite{Poulsen2021}. Its relation to the XFEL laboratory coordinate systems used in this work can be seen by comparing Figs.~\ref{fig:1} (a,b) and \ref{fig:S5} (a,e), and is given by

\begin{equation}
\mathbf{r}_{\ell, \textrm{XFEL}}  =  \begin{bmatrix}
          0 & 0 & 1 \\
          0 & -1 & 0\\
          1 & 0 & 0\\
         \end{bmatrix}\mathbf{r}_{\ell, \textrm{sync}}.
\end{equation}

In the geometrical optics formalism, several coordinate systems are used \cite{Poulsen2021}. The sample and the grain coordinate systems are defined in Fig.~\ref{fig:S5}. The sample coordinate system is rotated an angle $\theta$ relative to the laboratory coordinate system. The grain coordinate system is defined according to the crystallographic directions in the diamond crystal. An additional strain-wave coordinate system is defined to have $z_{\textrm{sw}}$ along the propagation direction of the strain wave, $[\bar{1}10]_{\textrm{g}}$, and $x_{\textrm{sw}}$ along $[001]_{\textrm{g}}$. 

Let us assume a \emph{longitudinal} wave propagating along $[\bar{1} 1 0]_{\textrm{g}}$. Given the 1D thermomechanical model the displacement gradient field, $\mathbf{F}(\vec{r},t)$ has only one non-trivial component: 

\begin{equation}
 \mathbf{F_{\textrm{sw}}} = \begin{bmatrix}
1 & 0 & 0\\
0 & 1 & 0\\
0 & 0 & 1+f(z_{\textrm{sw}})\\
\end{bmatrix},
\end{equation}
where $f(z_{\textrm{sw}})$ is the strain-profile in Fig.~\ref{fig:S4}.  $\mathbf{F}_{\textrm{sw}}$ is related to $\mathbf{F}_{\textrm{g}}$ by equation (75) in Ref.~\cite{Poulsen2021}:

\begin{equation}
 \mathbf{F_{\textrm{g}}} = \mathbf{U}_{\textrm{sw}} \mathbf{F_{\textrm{sw}}}\mathbf{U}^{\textrm{T}}_{\textrm{sw}},
\end{equation}
with the coordinate transform

\begin{equation}
\mathbf{r}_{\textrm{g}} =  \mathbf{U}_{\textrm{sw}}\mathbf{r}_{\textrm{sw}}.
\end{equation}

Figure \ref{fig:S5} shows the relation between the grain coordinate system and the strain wave coordinate system. $x_{\textrm{sw}}$ lies along $[001]_{\textrm{g}}$, $y_{\textrm{sw}}$ lies along $[110]_{\textrm{g}}$ and  $z_{\textrm{sw}}$ lies along $[\bar{1}10]_{\textrm{g}}$. The columns of $\mathbf{U}_{\textrm{sw}}$ are the basis vectors in the strain-wave coordinate system expanded in the basis of the grain coordinate system. Hence,

\begin{equation}
\mathbf{U}_{\textrm{sw}} =  \begin{bmatrix}
          0 & \frac{1}{\sqrt{2}} &  - \frac{1}{\sqrt{2}}\\
           0 & \frac{1}{\sqrt{2}} &   \frac{1}{\sqrt{2}}\\
          1&0 &0  \\
         \end{bmatrix}.
\end{equation}

The orientation of the diamond crystal $\mathbf{U}$ enters into the formalism by the definition of the "grain" system  Ref.~\cite{Poulsen2021}:

\begin{equation}
\mathbf{r}_{\textrm{s}} =  \mathbf{U}\mathbf{r}_{\textrm{g}}.
\end{equation}
In Fig.~\ref{fig:S5}, it can be seen that $x_{\textrm{s}}$ is parallel to $[110]_{\textrm{g}}$ $\times$  $[1\bar{1}\bar{1}]_{\textrm{g}}$ = [$\bar{1}1\bar{2}]_{\textrm{g}}$, $y_{\textrm{s}}$ is parallel to $[110]_{\textrm{g}}$ and  $z_{\textrm{s}}$ is parallel to $[1\bar{1}\bar{1}]_{\textrm{g}}$. The columns of $\mathbf{U}$ are the basis vectors in the grain coordinate system expanded in the basis of the sample coordinate system. Hence,

\begin{equation}
\mathbf{U} =  \begin{bmatrix}
           -\frac{1}{\sqrt{6}} & \frac{1}{\sqrt{6}} & -\frac{2}{\sqrt{6}} \\
            \frac{1}{\sqrt{2}} & \frac{1}{\sqrt{2}} & 0 \\
             \frac{1}{\sqrt{3}} & -\frac{1}{\sqrt{3}} & -\frac{1}{\sqrt{3}} \\
         \end{bmatrix}.
\end{equation}

Both $\mathbf{U}$ and $\mathbf{U}_{\textrm{sw}}$ satisfy $\mathbf{A}^{\textrm{T}} = \mathbf{A}^{-1}$ and det($\mathbf{A})$ =1.

\subsubsection*{Angular and Spatial resolution}

The geometrical optics formalism introduced above simulates the 6D reciprocal space-direct space resolution function \cite{Poulsen2021}. This was used to optimise the set up prior to actual beamtime \cite{Holstad2022}. For reference we here provide results for the parameters actually used during experiment.

\begin{figure}[ht!]
\centering
\includegraphics[width = 0.75\linewidth]{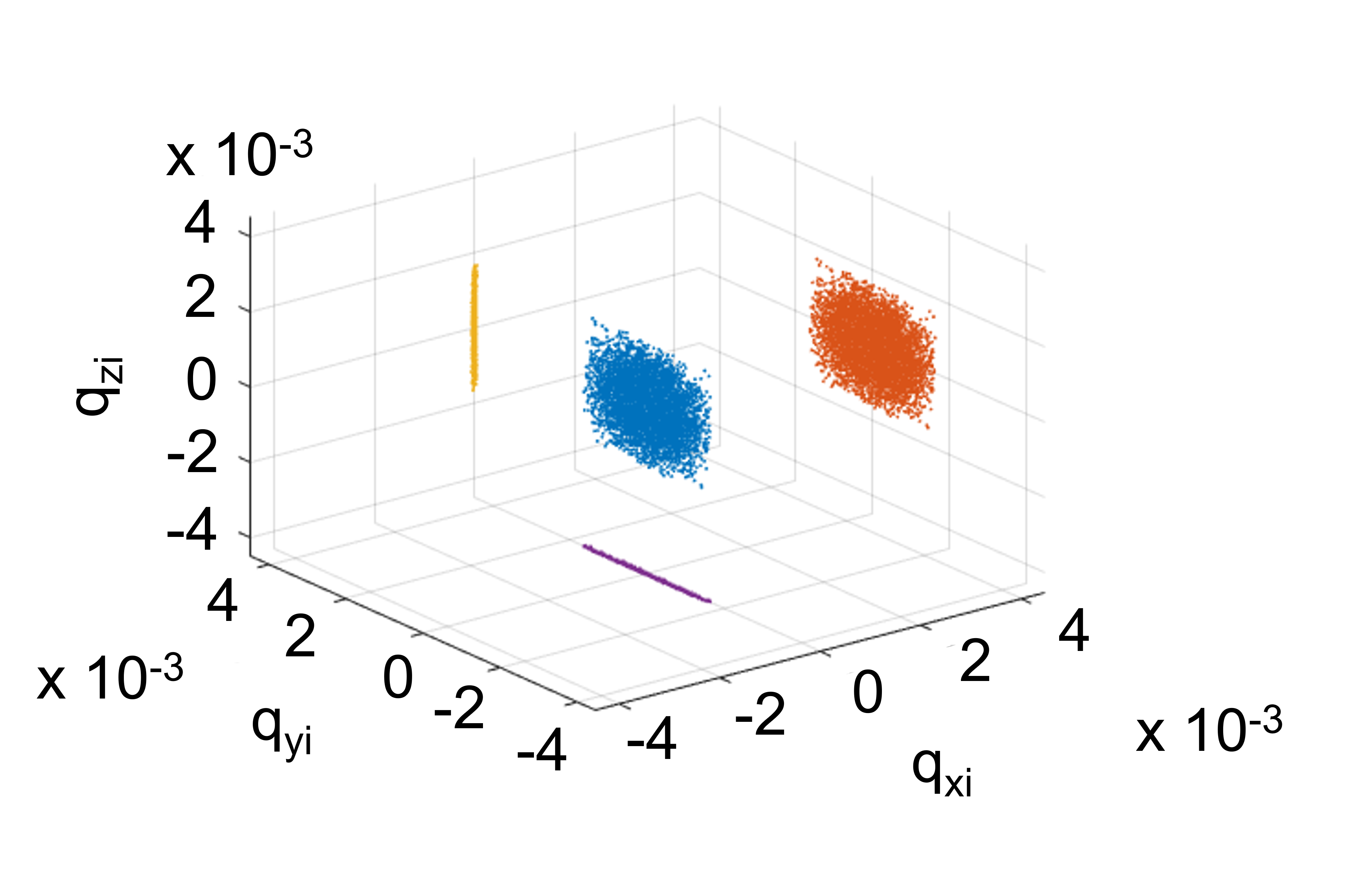}
\caption{Reciprocal space resolution function for the DFXM set up at LCLS. The simulation presented (for visualisation purposes) involved   10\,000 simulated rays. Dark blue cloud at center: 3D scatter plot for the resolution function expressed in the imaging coordinate system. The purple, orange and yellow symbols correspond to 2D projections onto the $q_{x_i}$-$q_{y_i}$ plane, $q_{y_i}$-$q_{z_i}$ plane and the $q_{x_i}$-$q_{z_i}$ plane, respectively. $(x_i,y_i,z_i)$ denotes the imaging coordinate system (see text, and Ref.~\cite{Poulsen2021}).}
\label{fig:S6}
\end{figure}

 \emph{Reciprocal space resolution function}. A Monte Carlo ray simulation of the reciprocal space resolution function is visualised in Fig. \ref{fig:S6}. (The imaging coordinate system is used, which corresponds to the laboratory coordinate system rotated by 2$\theta$ around $y_{\ell}$ \cite{Poulsen2021}). 10\,000 rays were used in Fig.~\ref{fig:S6} to visualize the anisotropy. Comparison of the projection shown in orange with those in yellow and purple shows a large anisotropy in the resolution function. To first order, the resolution function is a disc, with a ``thin dimension'' parallel to the optical axis of the objective.  The dimensions of the two wide axes are defined by the acceptance functions set by the numerical aperture of the objective, NA, producing a nearly planar distribution. The NA is larger than the maximum strain in the acoustic waves. In the actual simulations, 100 million rays were used for accuracy, filling 500 points in a range of 2.5E-3 along the three axes in reciprocal space. The energy band width (FWHM) was $\Delta E /E$ $\sim$ $10^{-4}$, the divergence (FWHM) $\Delta \zeta$ = 30 µrad in both horizontal and vertical directions. In the objective CRL, the lenslets had a radius of curvature of $R =$ 50 µm, and a center-to-center distance between successive lenslets $T = $ 1 mm. The sample-to-objective-entry plane distance was $d_1 = 0.23$ m. The NA was 3.598E-04 (root-mean-square).

The majority of the crystal is strain free and will therefore give rise to diffraction at the Origo with $\vec{Q} = \vec{Q}_0$. The sound wave may be visible in this “strong beam” but dynamical diffraction makes it difficult to quantify such images. For this reason, in this experiment "weak beam contrast" is applied, implemented by a rotation in $\phi$. The range of the strong beam condition is given by the width (FWHM) of the plate - this is according to the simulations $\Delta \phi$ $<$ $10^{-4}$, as shown in Fig.~\ref{fig:S10} (c).

\begin{figure}[!ht] 
\centering
  \centering
  \includegraphics[width=0.8\linewidth]{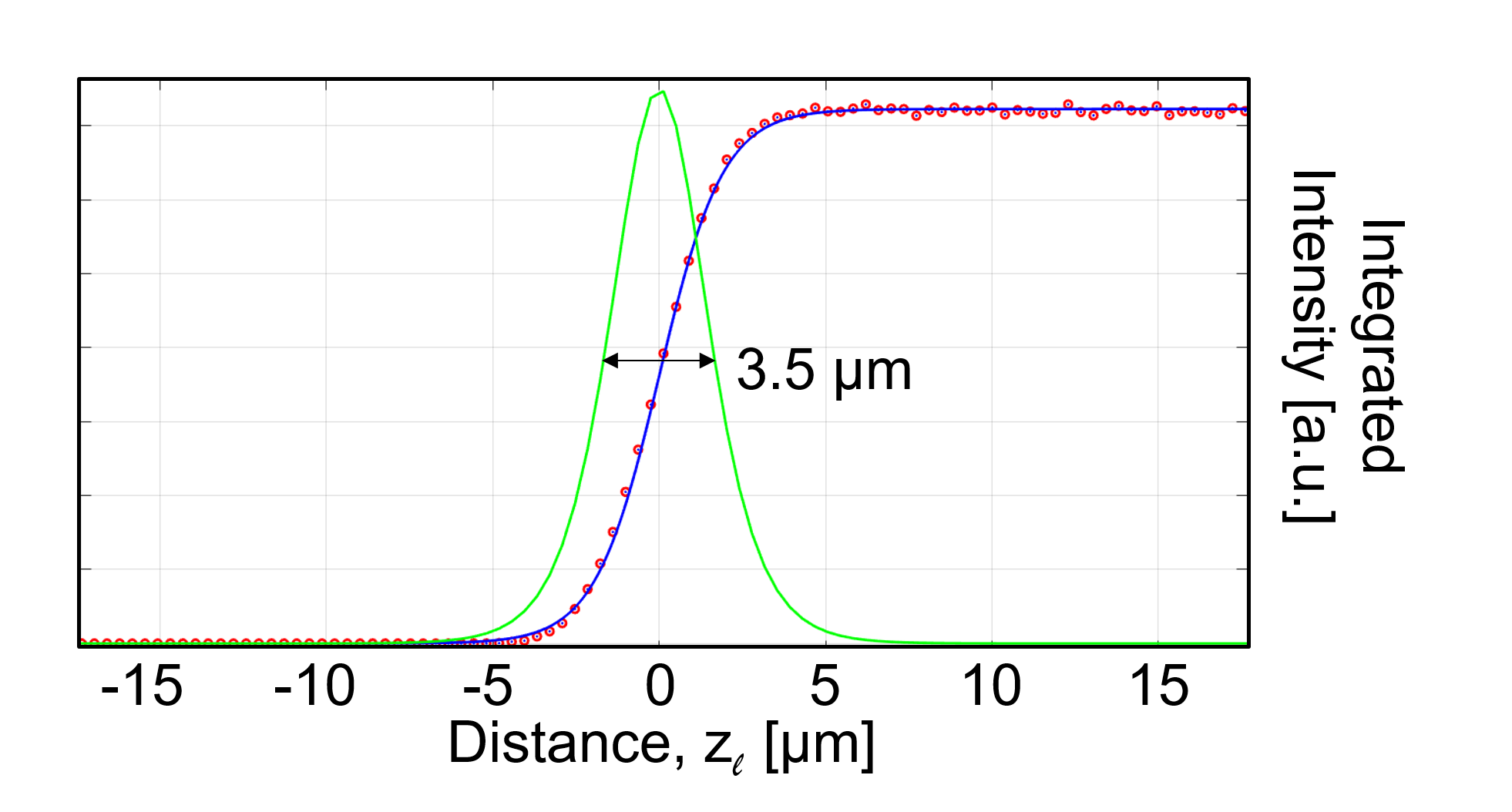}
  \caption{Simulated instrumental blur along $z_{\ell}$ for $\phi$ = +0.0764 mrad. Modelling the deformation gradient tensor field as a Heaviside function in $z_{\textrm{sw}}$ the geometrical optics forward simulator produces an intensity profile shown with red dots. A sigmoid function was fitted to this intensity profile (blue line). The FWHM of the derivative (green line) is 3.5 µm.}
  \label{fig:S7}
\end{figure}

\emph{Direct space resolution function}. 
The spatial resolution function is anisotropic, dominated by the beam width. To illustrate the effect, Fig.~\ref{fig:S7} shows the intensity-profile across a strain-wave (integrated over $y_{\ell}$) resulting from having a Heaviside step-function along $z_{\textrm{sw}}$ in the geometrical optics simulation. The offset in $\phi$ was set to +0.0764 mrad (the offset in $\phi$ is defined relative to the center of mass of the rocking curve in the bulk; see Fig.~\ref{fig:S9} (c) below). 

\begin{equation}
\label{eq:heavi}
\mathbf{F_{\textrm{sw}}}  = \begin{bmatrix}
1 & 0 & 0\\
0 & 1 & 0\\
0 & 0 & 1-A \cdot H(z_{\textrm{sw}})\\
\end{bmatrix},
 \end{equation}
where $A$ is the amplitude of the perturbation, here chosen to be 2E-4 to reflect Fig.~\ref{fig:S4}. In Fig.~\ref{fig:S7}, the negatively strained part of the crystal (for $z_{\textrm{sw}}$ > 0) comes into the Bragg condition for $\phi$ $>$ 0, an observation that will be important in the analysis of rocking scans below.

The derivative of the step-function represent the instrumental blurring associated with a strain wave propagating along $z_{\textrm{sw}}$. In the two orthogonal directions the spatial resolution is to a first approximation given by the effective pixel size within the sample, as the angular contributions arising from the divergence of the incident beam and the energy bandwidth are small. 

\subsubsection*{Comparison of intensity profiles at fixed offset in $\phi$}
\begin{figure}[t!] 
\centering
  \centering
  \includegraphics[width=.95\linewidth]{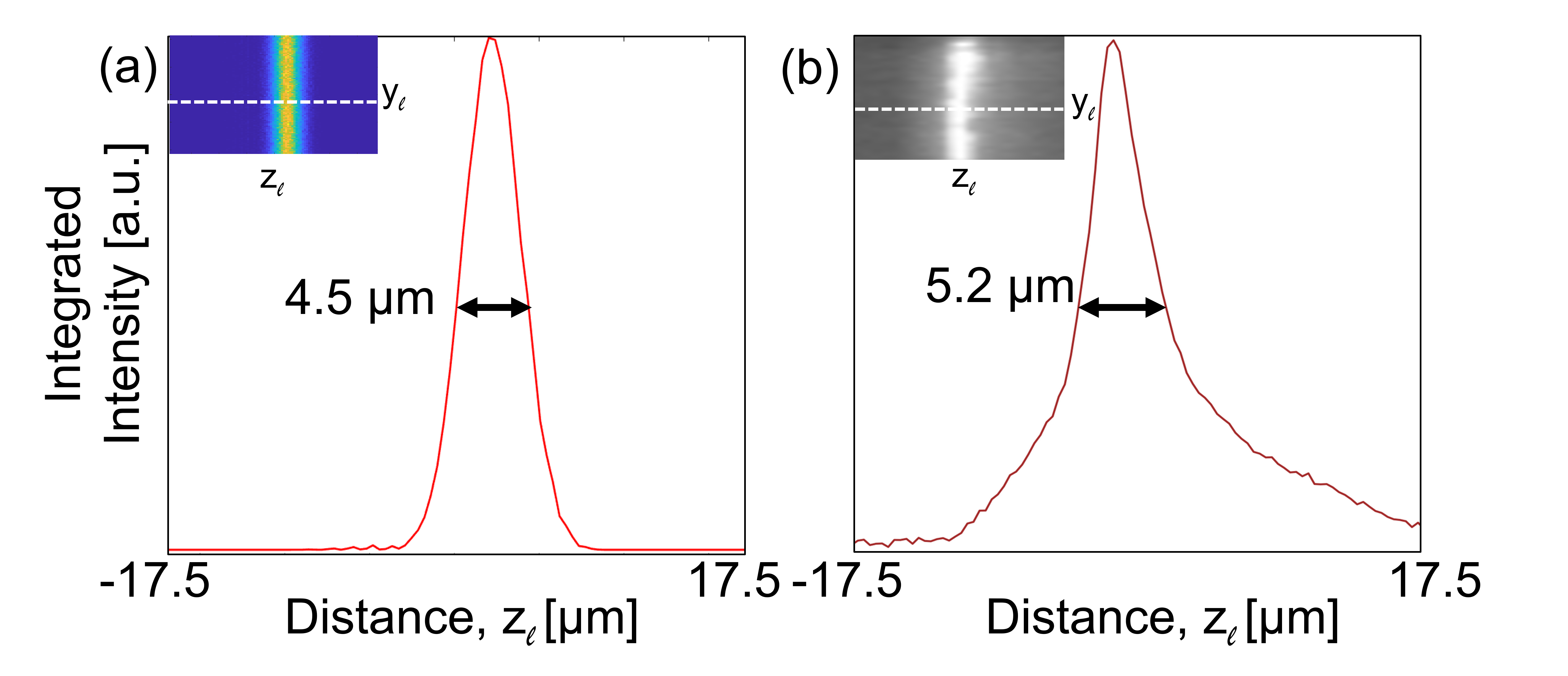}
  \caption{Comparison of profiles of the fast strain wave for an offset in $\phi$ from the strong beam condition by +0.0764 mrad. (a) As simulated by a combined thermomechanical and X-ray geometrical optics forward simulation \cite{Holstad2022}. The simulation relates to a planar longitudinal acoustic wave travelling in direction $[\bar{1} 1 0]_{\textrm{g}}$. The integrated line profile of the peak is shown.  The inset shows the full DFXM image with the $z_{\ell}$-axis marked by a dashed line. Here yellow signifies maximum intensity and dark blue minimum intensity. (b) Corresponding experimental DFXM data.}
  \label{fig:S8}
\end{figure}

Figure \ref{fig:S8} (a) shows a simulated DFXM image for an angular offset in $\phi$ of +0.0764 mrad (the offset in $\phi$ is defined relative to the center of mass of the rocking curve in the bulk; see Fig.~\ref{fig:S9} (c) below). A comparison of this and a line-out representing an integration in the vertical direction of the image with the experimental data in Fig.~\ref{fig:S8} (b) shows a satisfactory correspondence. The maximum intensity in the raw individual images is about 200 counts/pixel, cf. Movie 1. The maximum intensity in Fig.~\ref{fig:S8} (a) is $\sim$210 counts/pixel (using the noise model in Ref.~\cite{Holstad2022}, and assuming a third of the 1.6mJ/10.1keV = 9.9e11 photons make it past the monochromator). The strain wave in ~Fig.~\ref{fig:S8} (b) was taken from a position within the first 150 µm to avoid probing the curved part of the wave-front in Fig.~\textbf{\ref{fig:S1}}.

\subsubsection*{Comparison of rocking scans}

By varying $\phi$ we probe one shear strain component. The strain sensitivity is essentially given by the "thin" direction of the reciprocal space resolution function introduced in Fig. \ref{fig:S6}. 

\begin{figure}[h] 
\centering
  \centering
  \includegraphics[width=0.98\linewidth]{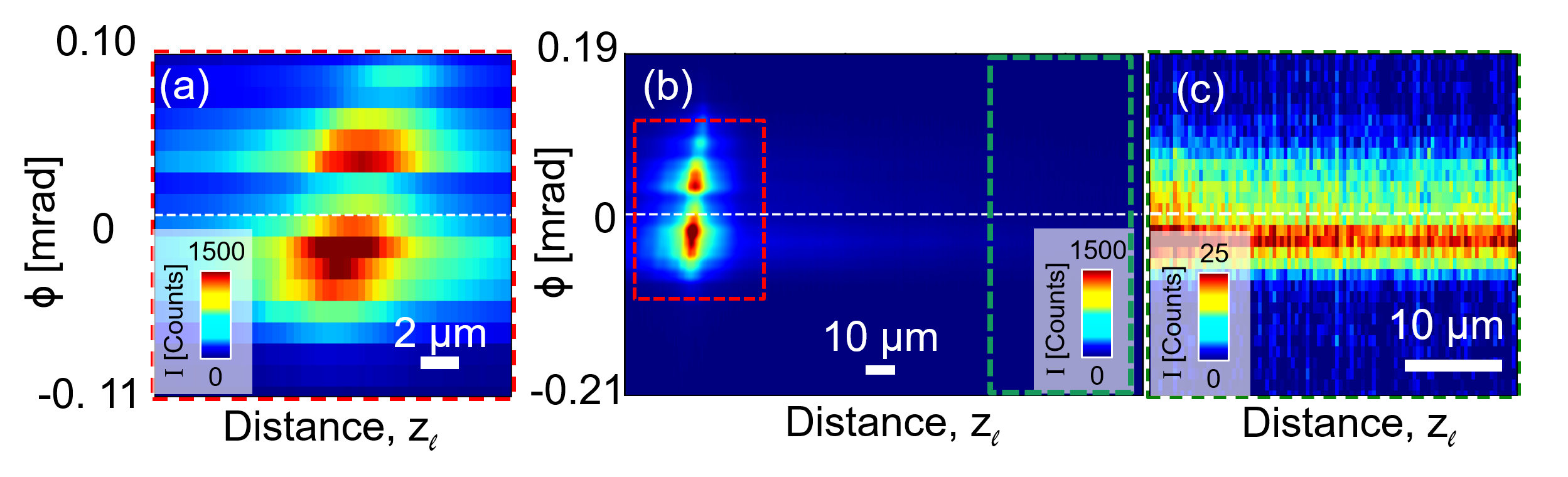}
  \caption{Experimental $\phi$-scan across a longitudinal strain-wave. (a) Zoom-in on the position of the strain-wave, showing two horizontally displaced strain-wave positions for different values of $\phi$. (b) Zoomed out overview. (c) Zoom-in on the bulk region marked with dashed green lines in (b), showing the bulk rocking-curve. The dashed white line describes the center of mass of the bulk rocking-curve, which we take as $\phi$ = 0.}
  \label{fig:S9}
\end{figure}

Experimentally a $\phi$ rocking-scan was performed at a delay of $\Delta t = 5.5$ ns, in steps of $\Delta \phi = 1.309 \cdot 10^{-5}$ rad. The resulting movie is summarised in Fig.~\ref{fig:S9}, displaying the intensity (averaged over $y_{\ell}$) as function of $z_{\ell}$ horizontally and $\phi$ vertically. The rocking curve at distances far from the strain wave is seen as representing the combined Darwin width and instrumental resolution function, and we set $\phi = 0$ at center of mass position (Fig.~\ref{fig:S9} (c)). In the vicinity of the fast strain wave, the distribution is asymmetric with two lobes, see Fig.~\ref{fig:S9} (a). The center of mass of these are separated by $\sim$1 µm along $z_{\ell}$ and by  and $\sim$0.05 mrad in $\phi$.

Notably the minimum intensity between these two is shifted by $1.963\cdot 10^{-5}$ rad in relation to $\phi = 0$, as defined by Fig.~\ref{fig:S9} (c). We speculate this is caused by a slight rotation of the lattice. We also note that the maximum intensity in sub-figure (a) is much higher than in (c). We attribute this to dynamical diffraction effects, to be explored in future studies.

\begin{figure}[h] 
\centering
  \includegraphics[width=1\linewidth]{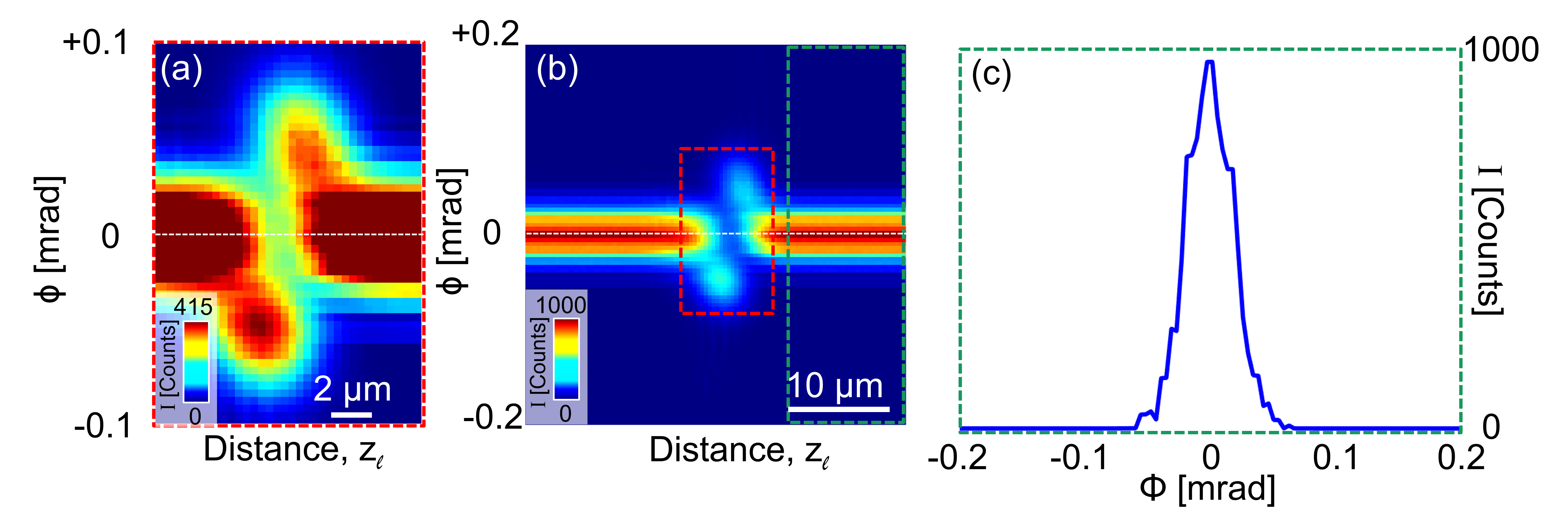}
  \caption{Geometrical optics simulated plot of $\phi$-scan across a longitudinal strain-wave. (a) Scaled to show the weak-beam. (b) Scaled to show the strong-beam. (c) Plot of rocking curve in bulk averaged over $z_{\ell}$ in the region marked with green dashed lines in (b). }
  \label{fig:S10}
\end{figure}

To further understand the data in Fig.~\ref{fig:S9}, we forward simulated DFXM images of the strain waves at different $\phi$ values (100 values in the range $\pm$ 0.2 mrad; $\Delta \phi = 4.04 \cdot 10^{-6}$ rad) using the 1D thermomechanical model to generate the strain waves in Fig.~\ref{fig:S4} and the geometrical optics simulation tools already described. From these simulations, we generated 2D plots similar to those of Fig.~\ref{fig:S9}. The result can be seen in Fig.~\ref{fig:S10}.

Figure~\ref{fig:S10} (a) is scaled to a maximum of 415 counts per pixel. Two lobes can be seen, with centers of mass that are displaced in a qualitatively similar fashion to the lobes in the experimental data. However, quantitatively the separation-distances between the centers of mass of $\sim$0.1 mrad in $\phi$, and $\sim$2 µm along $z_{\ell}$, are both significantly larger than the separation between the centers of mass in the experimental data in Fig.~\ref{fig:S9} (a). We speculate that this reflects effects of dynamical diffraction, uncertainties in the model parameters and/or inadequacies related to using a 1D thermomechanical strain-wave model for a 3D process (see Fig.~\textbf{\ref{fig:S1}}).

Figure~\ref{fig:S10} (b) is scaled to a maximum of 1000 counts per pixel, corresponding to the strong-beam condition. In the simulation, the position of  $\phi$ = 0 (defined by taking the center of mass of the rocking-curve in Fig.~\ref{fig:S10} (c)) is symmetric between the two strain-wave maxima.

\FloatBarrier
\subsubsection*{Smearing due to beam-thickness}
The aim of this section is to provide an intuitive understanding of the effect of the beam thickness on the strain-wave measurements. The FWHM thickness along $x_{\ell}$ is 3.9 µm. Projected onto the [$\bar{1}10]_{\textrm{g}}$ direction this corresponds to 3.9 $\cdot$ cos(52.78$^{\circ}$) µm $\approx$ 2.4 µm.  

\begin{figure}[h] 
\centering
  \centering
  \includegraphics[width=0.9\linewidth]{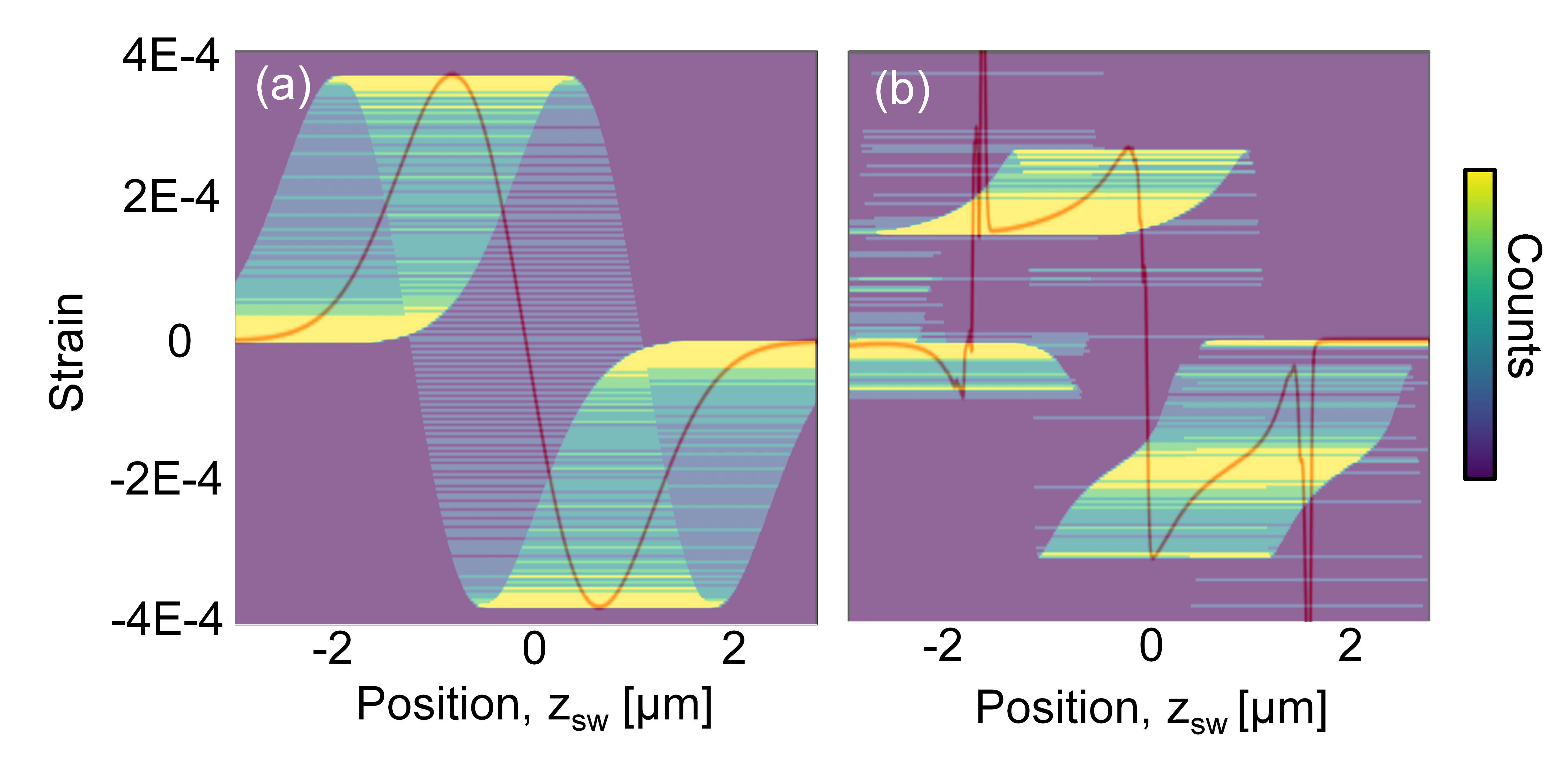}
  \caption{Effect of beam width on strain profile using as models for the longitudinal strain wave (red lines) (a) an idealized model, see text, and (b) a 1D thermomechanical simulation. In both cases the profiles are smeared out by a box-function of width 2.4 µm. The overlaid colormaps are the resulting intensity profiles.}
  \label{fig:S11}
\end{figure}

Assume a box shaped smearing function with this FWHM. In Fig.~\ref{fig:S11} (a) we plot a histogram of the resulting strain values as function of position for a simple strain wave model (red curve),

\begin{equation}
    f(z_{\textrm{sw}}) = -\frac{A}{s} \cdot (z_{\textrm{sw}}-z_0)\cdot e^{\frac{(z_{\textrm{sw}}-z_0)^2}{2s^2}},
\end{equation}
where $A$ is the amplitude of the strain wave, $s$ a parameter describing its spatial width, and $z_0$ the central position. 
The overlaid colormap is a plot of the histograms. Next, the same histogram representation is made based on the one-dimensional thermomechanical model in Fig.~\ref{fig:S4}, resulting in Fig. \ref{fig:S11} (b). 

Figure \ref{fig:S11} shows that when the strain-profile is convoluted with the beam-thickness, a given part of the strain-wave profile can be detected in multiple positions of the box-beam, and when the strain-profile becomes flat, e.g.~at the extreme positions, similar strain-values add up, leading to a maximum in the histograms. In the experimental data, this would manifest as maxima in the rocking-curve data, such as those seen in Figs.~\ref{fig:S9} (a) and ~\ref{fig:S10} (a) (when the comparison is made it must be remembered that negative (positive) strain along $[\bar{1}10]_{\textrm{g}}$ leads to the Bragg-condition shifting to $\phi$ > 0 ($\phi$ < 0), as noted in the discussion of Fig.~\ref{fig:S7}).

\end{document}